\documentclass[fleqn,usenatbib, letters]{mnras}

\usepackage{newtxtext,newtxmath}
\usepackage[T1]{fontenc}
\usepackage{ae,aecompl}

\usepackage{graphicx}
\usepackage{bm}		
\usepackage{float}
\usepackage[caption = false]{subfig}
\usepackage{pdflscape}
\usepackage[flushleft]{threeparttable}
\title[Bulk Comptonization in 4C+25.05]{Bulk Comptonization: new hints from the luminous blazar 4C+25.05}
	
\author[Kammoun et al.]{E. S. Kammoun,$^{1}$\thanks{E-mail: \href{mailto:ekammoun@sissa.it}{ekammoun@sissa.it}} E. Nardini,$^{2}$ G. Risaliti,$^{3,2}$ G. Ghisellini,$^{4}$ E. Behar,$^{5}$ and A. Celotti$^{1,4,6}$
\\
$^{1}$SISSA, via Bonomea 265, I-34135 Trieste, Italy\\
$^{2}$INAF - Osservatorio Astrofisico di Arcetri, Largo E. Fermi 5, I-50125 Firenze, Italy\\
$^{3}$Dipartimento di Fisica e Astronomia, Universit\`{a} di Firenze, via G. Sansone 1, 50019 Sesto Fiorentino (Firenze), Italy	\\
$^{4}$INAF - Osservatorio Astronomico di Brera, via Bianchi 46, I-23807 Merate, Italy\\
$^{5}$Department of Astronomy, University of Maryland, College Park, MD 20742-2421, USA\\
$^{6}$INFN - Sezione di Trieste, via Valerio 2, I-34127 Trieste, Italy\\
}
\date{Accepted XXX. Received YYY; in original form ZZZ}

\pubyear{2017}

\begin{document}

\label{firstpage}
\pagerange{\pageref{firstpage}--\pageref{lastpage}}

\maketitle

\begin{abstract}
Blazars are often characterized by a spectral break at soft X-rays, whose origin is still debated. While most sources show a flattening, some exhibit a blackbody-like soft excess with temperatures of the order of $\sim$0.1 keV, similar to low-luminosity, non-jetted Seyferts. Here we present the analysis of the simultaneous {\it XMM-Newton} and {\it NuSTAR} observation of the luminous FSRQ 4C+25.05 ($z=2.368$). The observed 0.3--30 keV spectrum is best described by the sum of a hard X-ray power law ($\Gamma = 1.38_{-0.03}^{+0.05}$) and a soft component, approximated by a blackbody with $kT_{\rm BB} = 0.66_{-0.04}^{+0.05}$ keV (rest frame). If the spectrum of 4C+25.05 is interpreted in the context of bulk Comptonization by {\it cold} electrons of broad-line region photons emitted in the direction of the jet, such an unusual temperature implies a bulk Lorentz factor of the jet of $\Gamma_{\rm bulk}\sim 11.7$. Bulk Comptonization is expected to be ubiquitous on physical grounds, yet no clear signature of it has been found so far, possibly due to its transient nature and the lack of high-quality, broad-band X-ray spectra.
\end{abstract}

\begin{keywords}
galaxies: active -- (galaxies:) quasars: general -- (galaxies:) quasars: individual: 4C+25.05 -- X-rays: galaxies -- X-rays: general
\end{keywords}
\section{Introduction}
\label{sec:intro}

According to the unified models of active galactic nuclei \citep{Urry95}, radio-loud quasars (RLQs) are characterized by a relativistic jet emitting a collimated non-thermal continuum. If the jet points along the line of sight of the observer, then the quasar is classified as a blazar.The broad-band emission of blazars is characterized by two humps. The first hump (in radio/sub-millimetre frequencies) is attributed to synchrotron, while the second one (in X-/$\gamma$-ray frequencies) is attributed to inverse Compton (IC) processes. The seed photons for the IC process can be either intrinsic to the jet, emitted through synchrotron at low frequencies (synchrotron self-Compton, SSC), or originated from the accretion disc and reprocessed by the broad line region (BLR) and/or the molecular torus (external Compton, EC). The latter is likely the main process in sources revealing a pronounced dominance of the high-frequency hump over the synchrotron one. This usually occurs in the most powerful blazars, i.e. flat-spectrum radio quasars (FSRQs).

The X-ray spectrum of blazars hardens with increasing luminosity. In fact, FSRQs typically show photon  indices $\Gamma \sim 1.3$--$1.5$ in the 2--10\,keV band \citep[e.g.][]{Wilkes92, Boller00, Derosa08, Eitan13}, which are flatter than usually observed in less luminous RLQs \citep[$\Gamma \sim 1.75$;][]{Sambruna99} or radio-quiet quasars \citep[$\Gamma \sim 1.9$;][]{Piconcelli05}. Several blazars reveal a flattening in their X-ray spectra at energies below $\sim 2$\,keV with respect to a higher-energy power law \citep[e.g.][]{Fabian98, PiconcelliL05}. The origin of this flattening has been associated with intrinsic cold or warm absorption \citep[$N_{\rm H} \sim 10^{22}\, \rm cm^{-2}$; e.g.][]{Worsley06} or a break of the continuum due to intrinsic curvature of the EC emission from the jet \citep[e.g.][]{Ghisellini07, Tavecchio08, Paliya16}. Some sources, instead, reveal the presence of an excess in emission at similar soft energy ranges. Various scenarios have been suggested in order to explain this feature such as an excess due to the contribution of the accretion disc emission to the soft X-rays  \citep{Sambruna06}, similar to the one seen in radio-quiet AGNs, or an increase in the contribution of the SSC component \citep{Kataoka08}. Bulk Comptonization (BC) emission has been proposed as an alternative explanation of spectral flattening and/or excess \citep[e.g.][]{Begelman87, Sikora94, Celotti07}. In this context, cold (i.e. non-relativistic) leptons, travelling with a bulk Lorentz factor $\Gamma_{\rm bulk}$, would interact with the photons produced by the accretion disc, and with those reprocessed (re-isotropized) in the BLR and/or scattered by free electrons external to the jet. The BC emission of disc and BLR photons would result in excess emission with respect to the power-law continuum, respectively emerging in the far ultraviolet (hence not accessible) and mid X-ray ranges. The latter component would correspond to a hump peaking at $\sim$3\,keV, accompanied by a flattening towards soft energies, which can mimic absorption. The two flavours of the soft X-ray spectral break (deficit or excess) have been usually treated as originated from different processes. \citet{Celotti07} modelled the flattening seen in the blazar GB\,B1428+217 ($z=4.72$) assuming a transient BC scenario. The flattening in this source had been suggested to be due to the presence of intrinsic absorption with column densities exceeding $10^{22}\,{\rm cm^{-2}}$ \citep[e.g.][]{Boller00, Fabian01, Worsley06}. It should be noted that BC is expected to be present in all blazars but has never been confirmed, to our knowledge. Besides the case of GB\,B1428+217, \cite{Kataoka08} and \cite{Derosa08} presented tentative hints of the presence of BC in the FSRQs PKS\,1510--089 ($z=0.361$) and 4C+04.42 ($z=0.965$), respectively.

In this letter, we study the X-ray spectrum provided by a coordinated {\it XMM-Newton} and {\it NuSTAR} observation of the FSRQ 4C+25.05 \citep[a.k.a. PKS\,0123+257, $z = 2.368$, $\log (M_{\rm BH}/M_\odot)= 9.24 \pm 0.44$;][]{Kelly07}. The following cosmological parameters are assumed: $\Omega_{\rm M} = 0.27$, $\Omega_{\Lambda} = 0.73$, and $H_0 = 70\,{\rm km\,s^{-1}\,Mpc^{-1}}$.
\section{Observations and data reduction}
\label{sec:reduction}
4C+25.05 was observed  simultaneously by {\it XMM-Newton} and {\it NuSTAR}, on 2017 January 15 (Obs. IDs 0790820101 and 60201047002, respectively). The log of the observations is presented in Table\,\ref{table:log}. 

\begin{table}
\centering
\caption{Net exposure time, average net count rate and ratio of the source to total counts, in the observed 0.3--10\,keV band for EPIC-pn and MOS, and 3--30\,keV band for FPMA/B .}
\begin{tabular}{lccc}
\hline\\[-0.4cm]
Instrument	&	Net exposure	&			Count Rate 	& Source/total\\
 & (ks) & (Count\,s$^{-1}$) & \\ \hline \\[-0.4cm]
pn	&	37.7	&	$	0.522	\pm	0.003	$	&	99\%	\\
MOS	&	49.4	&	$	0.305	\pm	0.003	$	&	98.8\%	\\
FPMA	&	40.4	&	$	0.047	\pm	0.001	$	&	90.3\%	\\
FPMB	&	40.4	&	$	0.043	\pm	0.001	$	&	86.9\%	\\\hline
\end{tabular}
\label{table:log}
\end{table}

\begin{figure}
\centering
\includegraphics[width = 0.45\textwidth]{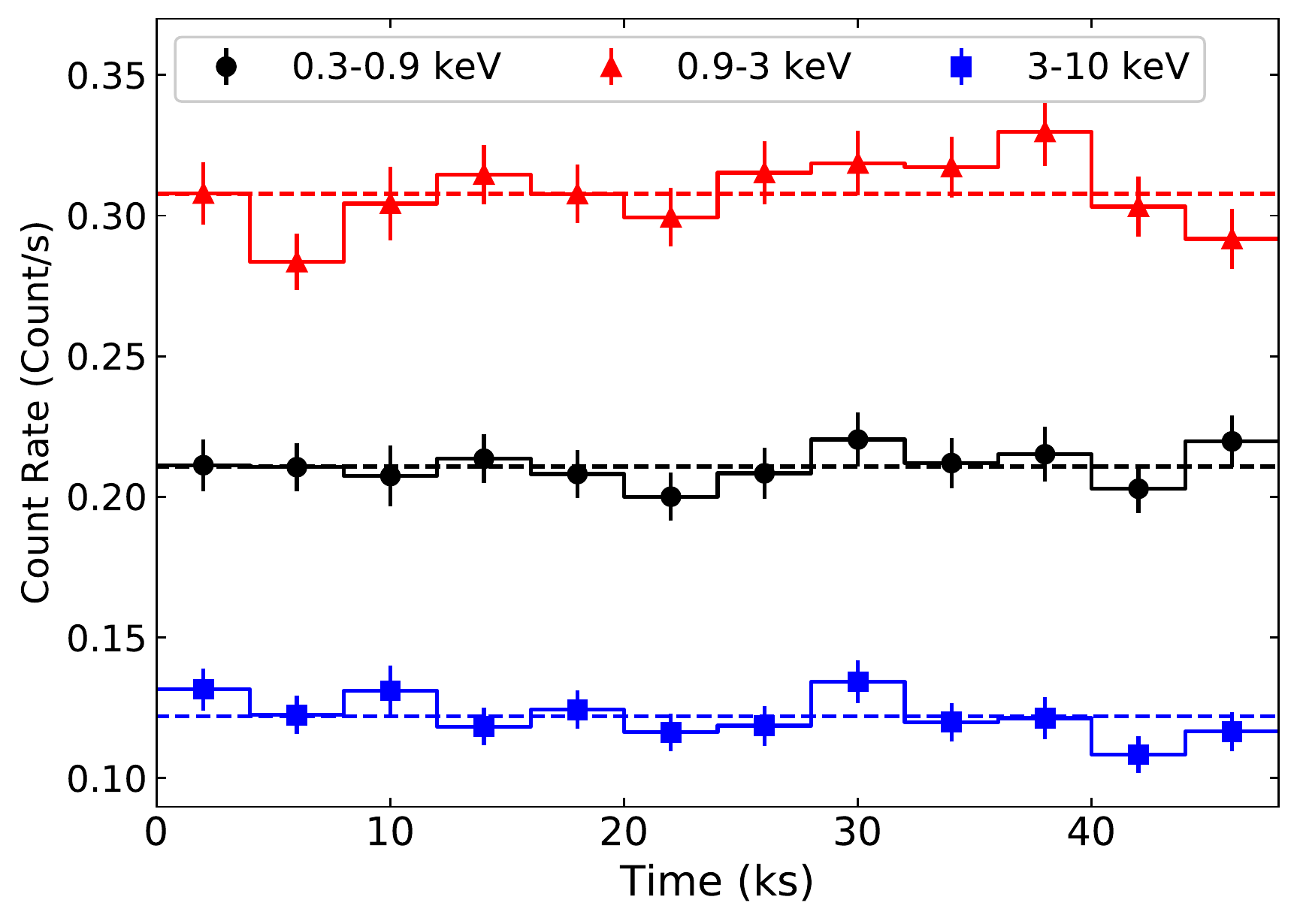}
\caption{{\it XMM-Newton} light curves (with a time bin of 4\,ks) extracted in the 0.3--0.9\,keV (black circles), 0.9--3\,keV (red triangles) and 3--10\,keV (blue squares) observed energy bands, corresponding to the $\sim$\,1--3\,keV, 3--10\,keV and 10--34\,keV rest-frame energy bands, respectively. The horizontal dashed lines correspond to the average count rate for each energy band.}
\label{fig:xraylc}
\end{figure}

We reduced the  {\it XMM-Newton} data using {\tt \small SA}S\,v.15.0.0 and the latest calibration files. We followed the standard procedure for reducing the data of the EPIC-pn \citep{Stru01} and the two EPIC-MOS \citep{Tur01} CCD cameras, all operating in full frame mode with a thin filter. The pn and MOS data were processed using {\tt EPPROC} and {\tt EMPROC}, respectively. Source spectra and light curves were extracted from a circular region of radius of 25\,arcsec for both instruments. The corresponding background spectra and lightcurves were extracted from an off-source circular region located on the same CCD chip, with a radius approximately twice that of the source. We filtered out periods with strong background flares (two minor ones happened in the middle and a major one towards the end of the observation), for a total of around 6\,ks. The light curves, background-corrected using {\tt EPICLCCOR}, are shown in Fig.\,\ref{fig:xraylc} in three different bands (soft, medium, hard). None of them shows any significant variability, as they are all well-fitted with a constant. Response matrices were produced using the {\tt FTOOLS} {\tt RMFGEN} and {\tt ARFGEN}. We rebinned the observed 0.3--10 keV spectra using the SAS task {\tt SPECGROUP} to have a minimum S/N of 5 in each energy bin. The MOS1 and MOS2 spectra are consistent, so we combined them using the {\tt SAS} command {\tt COMBINE}.


The {\it NuSTAR} data were reduced following the standard pipeline in the {\it NuSTAR} Data Analysis Software ({\tt NUSTARDAS}\,v1.6.0), and using the latest calibration files. We cleaned the unfiltered event files with the standard depth correction. We reprocessed the data with the {\tt saamode = optimized} and {\tt tentacle = yes} criteria for a more conservative treatment of the high background levels in proximity of the South Atlantic Anomaly. We extracted the source and background spectra from circular regions of radii 60 and 90\,arcsec, respectively, for both focal plane modules (FPMA and FPMB) using the {\tt HEASOFT} task {\tt NUPRODUCT}, and requiring a minimum S/N of 5 per energy bin. The spectra extracted from both modules are consistent with each other, so they are analysed jointly (but not combined together). Since the background starts to dominate above $\sim$30\,keV, we decided to analyse the {\it NuSTAR} data in the 3--30\,keV observed energy range, which corresponds to $\sim$10--100\,keV in the rest frame of the source. 

\section{Spectral analysis}
\label{sec:Xspec}

\begin{figure}
\centering
\includegraphics[width = 0.43\textwidth]{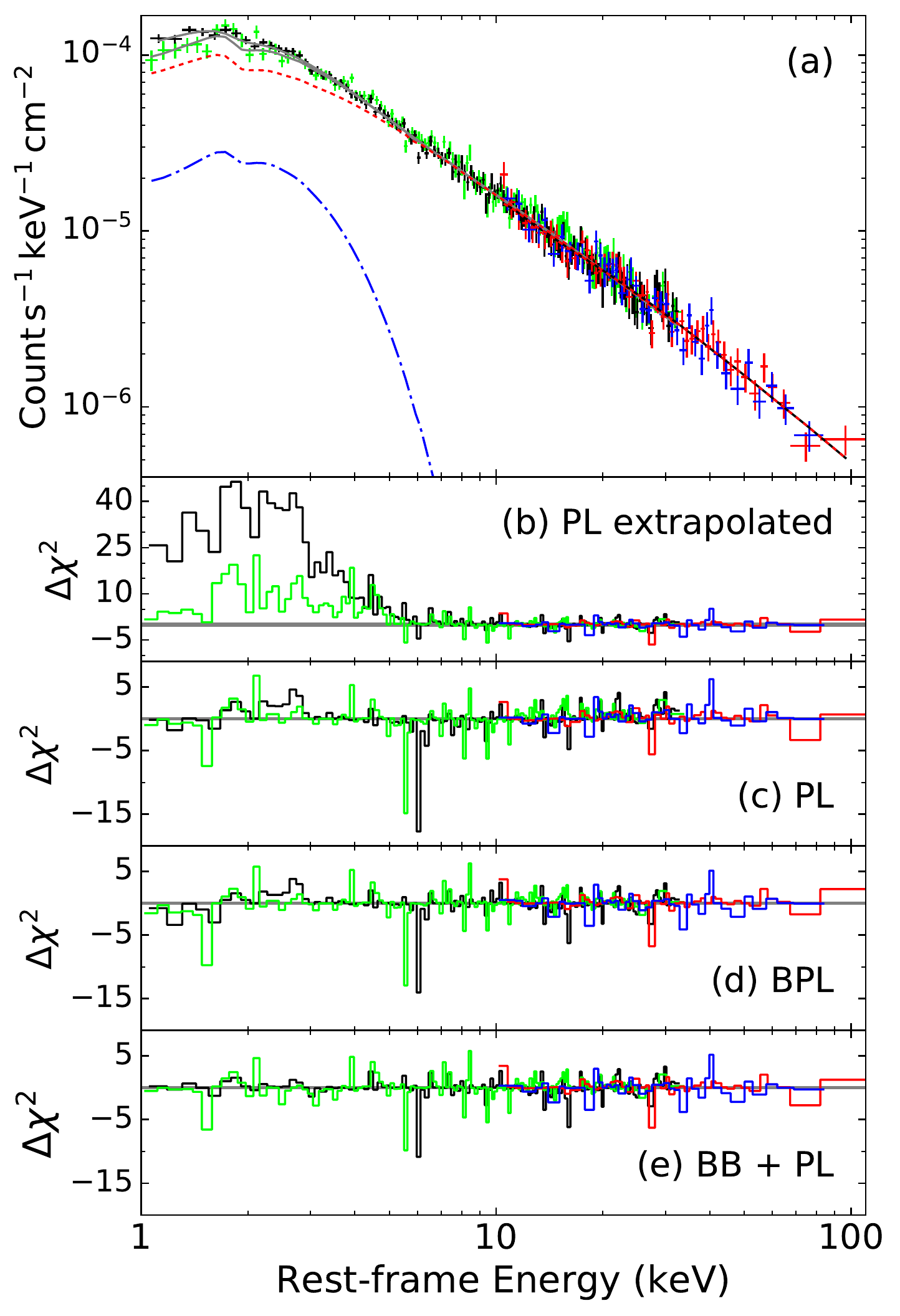}
\vspace*{-5pt}
\caption{Panel (a): Spectra from EPIC-pn (black), EPIC MOS (green) and {\it NuSTAR} FPMA/FPMB (red/blue) plotted together with the total best-fitting model (grey solid line) composed of a PL (red dotted line) and a BB component (blue dash-dotted line) plotted in the rest-frame energy. Panel (b): residuals obtained by fitting the spectra above 10\,keV (rest-frame) with a PL model then extrapolating it to softer energies, showing an excess in the $\sim$1--5\,keV range. Panels (c--e): residuals obtained by fitting the 1--100\,keV range with a PL, BPL, and BB+PL models, respectively (see \S\,\ref{sec:Xspec} for details).}
\label{fig:spectra}
\end{figure}

Throughout this work, spectral fitting was performed using {\tt XSPEC}\,v12.9s \citep{Arnaud96}. Unless stated otherwise, uncertainties on the parameters are listed at the 90 per cent confidence level ($\Delta \chi^2= 2.71$). We included a multiplicative constant, for each instrument, in order to account for the residual uncertainties in the flux calibration between the various detectors, fixing the constant for the EPIC-pn data to unity. The spectra from the various instruments are shown in Fig.\,\ref{fig:spectra}a.

We started by fitting the hard-energy part of the spectra above 3\,keV (observed) with a power-law model with a high-energy cutoff (hereafter PL model), taking into consideration the Galactic absorption in the line of sight \citep[$N_{\rm H} = 6.87\times10^{20}\, {\rm cm ^{-2}}$;][]{Kal05}. The fit is statistically acceptable ($\rm \chi^2/dof = 181/220$), showing no systematical trend in the residuals, with a photon index $\Gamma = 1.34 \pm 0.13$. We could only set a $3\sigma$ lower limit for the high-energy cutoff of $E_{\rm cut} > 63$\,keV (in the rest frame). The cross-calibration factor between {\it NuSTAR} and {\it XMM-Newton} was found to be $f_{\rm NuSTAR} =1.09 \pm 0.06$, consistent with the values reported by \cite{Madsen15}. The extrapolation of this model to lower energies reveals an excess below $\sim$5\,keV (rest-frame), as shown in Fig.\,\ref{fig:spectra}b. We re-fitted the same model to the full 0.3--30\,keV observed range. The fit is statistically acceptable ($\rm \chi^2/dof = 373/359$, $p_{\rm null}=0.29$). The residuals are not statistically significant, yet, qualitatively, they show a subtle overall curvature, leading to a systematic excess in the $\sim$1.5--3\,keV range. The best-fit photon index is steeper than the previous case ($\Gamma = 1.54 \pm 0.01$). $E_{\rm cut}$ is pegged to its maximum allowed value, with a $3\sigma$ lower limit of 387\,keV (in the rest frame). The cross-calibration factor between {\it NuSTAR} and {\it XMM-Newton} becomes larger, $f_{\rm NuSTAR} = 1.14 \pm 0.04$. We re-fitted the spectra accounting for the possible intrinsic neutral absorption in the rest frame of the source \citep[{\tt zTBabs;}\,][]{Wilms00}. The fit did not show any improvement ($\rm \chi^2/dof = 373/358$), revealing an intrinsic absorption that is consistent with zero (we could set a 3$\sigma$ upper limit on $N_{\rm H}$ to be $9.5\times 10^{20}\,\rm cm^{-2}$). We show the residuals corresponding to this model in Fig.\,\ref{fig:spectra}c. 

We considered several models, in order to account for the possible excess at soft energies. We first fitted the spectra with a broken power-law model (hereafter BPL model) modified by Galactic absorption only. The fit was statistically acceptable ($\rm \chi^2/dof = 343/357$) with the photon indices being $\Gamma_{\rm soft} = 1.56 \pm 0.03$, $\Gamma_{\rm hard} = 1.29 \pm 0.11$. The break and cutoff energies are $E_{\rm b} = 9.2_{-1.6}^{+1.2}$\,keV and $E_{\rm cut} = 121_{-44}^{+139}$\,keV in the rest frame. The 3$\sigma$ confidence level gives only a lower limit on $E_{\rm cut} > 60$\,keV. The cross-calibration factor is $f_{\rm NuSTAR} =1.08 \pm 0.05$. The fit improves by $\Delta \chi^2 = -30$ for one extra free parameter with respect to the PL model. This improvement is mainly due to the steepening of the spectrum at soft energies, leading to less prominent residual structures (as shown in Fig.\,\ref{fig:spectra}d), contrary to the commonly-seen flattening in other sources \citep[e.g.][]{PiconcelliL05}. 
\begin{table}
\centering
\caption{Best-fit parameters obtained by fitting the spectra with the PL, BPL and BB+PL models considered in this analysis. $E_{\rm b}$, $E_{\rm cut}$ and $kT_{\rm BB}$ are reported in the rest frame of the source. The last column represents the peak values of the 1D probability distribution obtained from the MCMC analysis.}
\begin{threeparttable}
\begin{tabular}{lcccc}
\hline\\[-0.4cm]
Parameter	&		PL				&		BPL				&		\multicolumn{2}{c}{BB+PL}						\\ \hline \\[-0.4cm]
$N_{\rm H}\, (10^{20}\rm cm^{-2})$	&	$	< 9.5^\dagger			$	&		--				&		--				&	--	\\[0.05cm]
$\Gamma_{\rm soft}$	&		--				&	$	1.56	\pm	0.03	$	&		--				&	--	\\[0.05cm]
$E_{\rm b}$\,(keV)	&		--				&	$	9.2	_{-1.6}	^{+1.2}	$	&		--				&	--	\\[0.05cm]
$\Gamma_{\rm hard}$	&	$	1.54	\pm	0.01	$	&	$	1.29	\pm	0.11	$	&	$	1.38	_{-0.03}	^{+0.05}	$	&	1.41	\\[0.05cm]
$E_{\rm cut}$\,(keV)	&	$	>387^\dagger			$	&	$	121	_{-44}	^{+139}	$	&	$	205	_{-54}	^{+256}	$	&	210	\\[0.05cm]
$N_{\rm PL} \times 10^{-4}$	&	$	3.05	\pm	0.05	$	&	$	3.13	\pm	0.07	$	&	$	2.63	_{-0.08}	^{+0.11}	$	&	2.65	\\[0.05cm]
$kT_{\rm BB}$\,(keV)	&		--				&		--				&	$	0.66	_{-0.04}	^{+0.05}	$	&	0.66	\\[0.05cm]
$N_{\rm BB} \times 10^{-5}$	&		--				&		--				&	$	1.92	_{-0.51}	^{+0.32}	$	&	1.8	\\ \hline\\[-0.4cm]
$\chi^2/{\rm dof}$	&		373/358				&		343/357				&		322/357				&		\\ \hline 
\end{tabular}
\begin{tablenotes}
	\item[$\dagger$] 3$\sigma$ lower/upper limit.
\end{tablenotes}
\label{table:param}
\end{threeparttable}
\end{table}


We also tested a reflection model \citep[{\tt RELXILL;}\,][]{Dauser13, Dauser16}, even though we do not detect any clear presence of either an iron line or a Compton hump. We fixed the spin to its maximum value (0.998) and we assumed a power-law illumination profile ($q = 3$) with a reflection fraction equal to unity. First, we let the inclination free to vary. The fit resulted in a very high inclination of the disc, which is not physical for this system. Then we fixed the inclination to 5$^\circ$, which corresponds to a nearly face-on configuration. The fit is statistically acceptable ($\rm \chi^2/dof = 362/357$). The best-fit photon index, ionization parameter and iron abundance are: $\Gamma = 1.43_{-0.08}^{+0.04}$, $\log \xi = 3.72_{-0.47}^{+0.08}$, and $A_{\rm Fe} = 3.7 \pm 1.2$\,solar, respectively. However, this model cannot be considered as a plausible explanation for the spectrum of 4C+25.05. On the one hand, it gives a statistically worse fit compared to the (phenomenological) BPL model. On the other hand, the Doppler-boosted featureless jet emission is expected to be much stronger than any reflection component, thus diluting all other features. Furthermore, it should be noted that it is unlikely in blazars to have a standard (i.e. similar to the non-jetted sources) X-ray corona. If the X-ray source illuminating the disc is a relativistically outflowing corona, then beaming effects would reduce the illumination of the disc, so any reflection component would be negligible \citep[see also][]{King17}.

\begin{figure}
\centering
{\includegraphics[width = 0.45\textwidth]{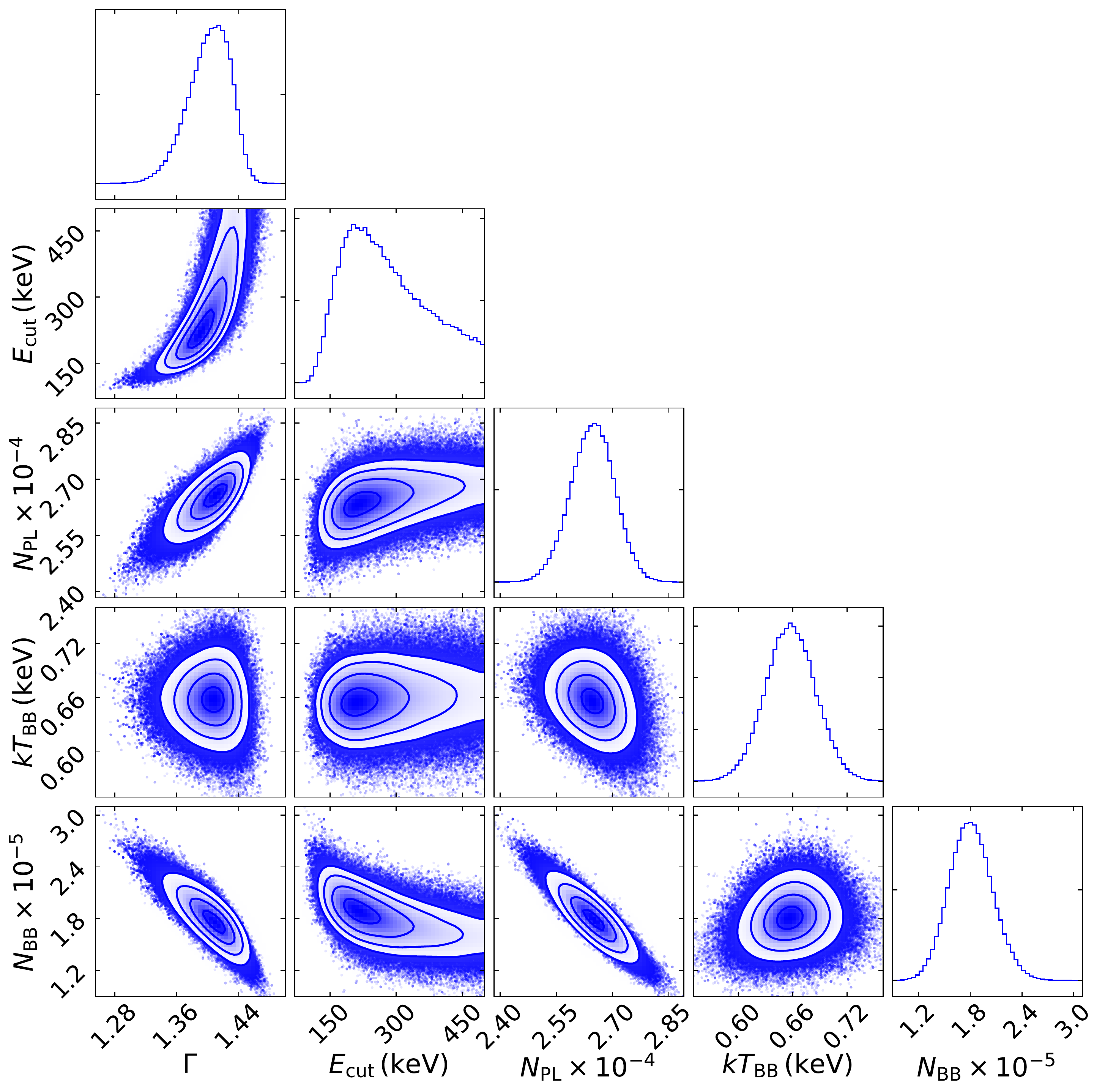}}
\vspace*{-5pt}
\caption{Results of the MCMC analysis of the PL+BB model. We show the outputs for photon index $\Gamma$, cutoff energy $E_{\rm cut}\,\rm(keV)$, blackbody temperature $kT_{\rm BB}\,({\rm keV})$, and normalizations of the power-law and the blackbody components $N_{\rm PL}$ and $N_{\rm BB}$, respectively. The 1D histograms show the probability distribution for each parameter normalized to unity.}
\label{fig:contour}
\end{figure}

Motivated by the BC model \citep{Celotti07}, we considered a model defined as the sum of a blackbody component and a power law (hereafter BB+PL model). The model fits very well the data ($\chi^2/{\rm dof}= 322/357$), without any systematic residuals. The best-fitting BB+PL model and the corresponding residuals are presented in Fig.\,\ref{fig:spectra}a,e, respectively. The best-fitting parameters of the BB+PL model, and, for comparison, of the PL and BPL models as well, are presented in Table\,\ref{table:param}. The errors on the parameters, for the BB+PL model, are calculated from a Markov chain Monte Carlo (MCMC)\footnote{We use the {\tt XSPEC\_EMCEE} implementation of the {\tt PYTHON EMCEE} package for X-ray spectral fitting in XSPEC by Jeremy Sanders (\url{http://github.com/jeremysanders/xspec_emcee}).} analysis, starting from the best-fitting model that we obtained. We used the Goodman-Weare algorithm \citep{Goodman10} with a chain of $500,000$ elements (100 walkers and 5000 iterations), and discarding the first $75,000$ elements as part of the `burn-in' period. The rest-frame temperature of the blackbody is $kT_{\rm BB} =0.66_{-0.04}^{+0.05}$\,keV, equivalent to a peak energy $E_{\rm peak,\, BB} = 3.93 kT_{\rm BB} \simeq 2.6 \pm 0.2$\, keV (in $\nu F_\nu$), consistent with the expected value \citep[e.g.][]{Celotti07}. In this case, the values of the photon index and the high-energy cutoff are $\Gamma = 1.38_{-0.03}^{+0.05}$ and $E_{\rm cut} = 205_{-54}^{+256}$\,keV (rest frame), respectively. These values are consistent with the ones that we obtained by fitting a PL model to the spectra above 3\,keV, and the hard component of the BPL model. The 3$\sigma$ confidence level gives only a lower limit on $E_{\rm cut} > 110$\,keV. The results of the MCMC analysis are shown in Fig.\,\ref{fig:contour}. We note that the peak values of the 1D probability distribution obtained from the MCMC analysis (presented in the last column of Table\,\ref{table:param}) do not exactly coincide with the best-fit values obtained by minimizing the $\chi^2$ value, but they are consistent within 1$\sigma$. The contours show that the temperature of the blackbody is not degenerate with any other parameter, confirming the presence of a significant excess over this energy range. Instead, we found some degeneracy between the normalizations of the PL and that of the BB and between both normalizations and the photon index. The mild degeneracy between the BB normalization with the photon index indicates that a steepening in the PL slope tends to compensate for the BB component. Furthermore, we determine the flux of the source to be $F_{0.3-10} = 2.25_{-0.05}^{+0.01}\times 10^{-12}\,{\rm erg\,s^{-1}\, cm^{-2}}$ in the 0.3--10\,keV observed energy range, which corresponds to a luminosity of $L_{1-30}=(9.44\pm 0.07)\times 10^{46}\,{\rm erg\,s^{-1}}$ in the 1--30\,keV rest-frame energy range. The errors on the flux and luminosity represent the 1$\sigma$ confidence level.

\section{Discussion and conclusions}
\label{sec:discussion}
\begin{figure}
\centering
\includegraphics[width = 0.45\textwidth]{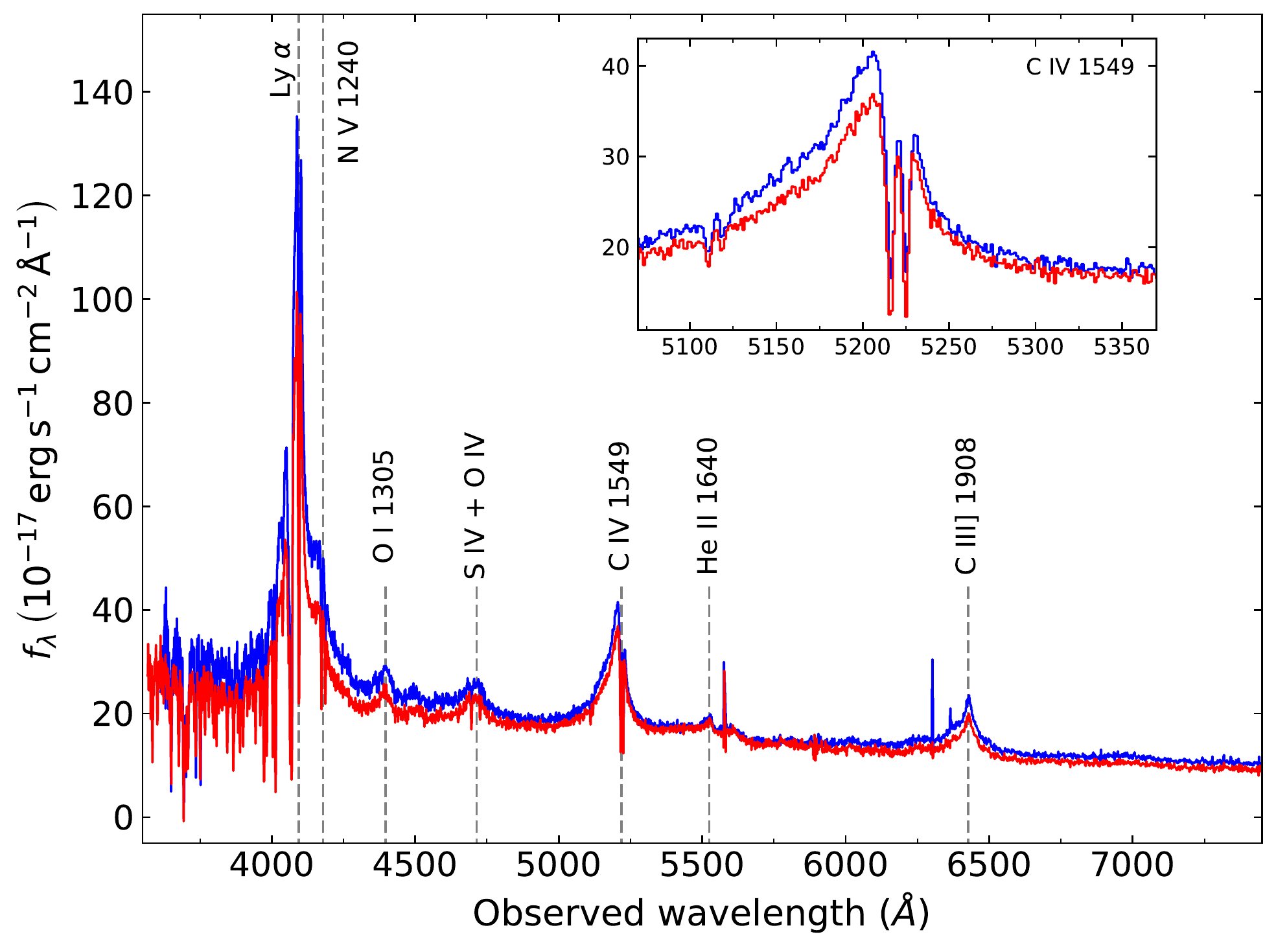}
\vspace*{-5pt}
\caption{Optical spectra of 4C+25.05 obtained from the SDSS-IV. The inset shows the peak of the \ion{C}{iv} emission line and the absorption doublet.}
\label{fig:SDSS}
\end{figure}

We have presented the X-ray spectral analysis of the simultaneous {\it XMM-Newton} and {\it NuSTAR} observation of the FSRQ 4C+25.05 at $z=2.368$. The observed spectrum of the source in the 0.3--30\,keV range can be well explained by (Table\,\ref{table:param}):

\begin{enumerate}

	\item[(i)] A simple power-law model ($\rm \chi^2/dof = 373/358$) of photon index 
	$\Gamma = 1.54$. Only upper and lower limits can be set for any intrinsic absorption 
	and high-energy cutoff, respectively. 
	
	\item[(ii)] A broken power-law model ($\rm \chi^2/dof = 343/357$), with 
	$\Gamma_{\rm soft} = 1.56$, $\Gamma_{\rm hard} = 1.29$, break energy of $E_{\rm b} 
	= 9.2$\,keV and cutoff at $E_{\rm cut} = 121$\,keV (rest frame). 	
	
	\item[(iii)] A power-law plus blackbody model ($\rm \chi^2/dof = 322/357$), 
	with $\Gamma = 1.38$, $E_{\rm cut} = 205$\,keV and $kT_{\rm BB} = 0.66$\,keV.

\end{enumerate} 
\noindent
The BPL model is largely phenomenological, yet the steeper $\Gamma_{\rm soft}$ with respect to $\Gamma_{\rm hard}$ clearly indicates a softening in the soft X-rays, as opposed to the flattening (due to either absorption or intrinsic curvature of the EC emission from the jet) usually observed in high-$z$ blazars. This excess favours the possibility that we are witnessing the spectral signature of bulk Comptonization of BLR photons. Indeed, the BB+PL model has the lowest $\chi^2$ among the various models that we consider in this work. The absence of any absorption in 4C+25.05 is confirmed by the two optical spectra obtained from the Sloan Digital Sky Survey \citep[SDSS-IV;][]{eBOSS,SDSS} in 2012 and 2015 (red and blue lines in Fig.\,\ref{fig:SDSS}, respectively), which do not show any signature of intrinsic cold absorption. We note that the SDSS spectra reveal the presence of narrow `associated' absorption lines from Ly$\alpha$,  \ion{N}{v}, \ion{Si}{iv} and \ion{C}{iv} (the \ion{C}{iv}\,$\lambda\lambda 1548,1550$ doublet is shown in the inset in Fig.\,\ref{fig:SDSS}). These features, rather common in flat-spectrum quasars \citep[e.g.][]{Richards99, Richards01, Baker02}, indicate the presence of an ionized low-density ($N_{\rm \ion{H}{i}} \lesssim 10^{17}\,\rm cm^{-2}$) absorber close to the central engine. A detailed discussion of these narrow features and of their origin goes beyond the scope of this work. However, we note that this absorption system has been suggested to partially cover the background light (including both the continuum and broad emission-line region) by \cite{Barlow97}. We finally point out that the SDSS spectra reveal several broad emission lines, implying the presence of a proper BLR in this source. The BLR component has been already detected and well-discussed in previous studies \citep[e.g.][]{Baldwin77, Padovani89, Kelly07}.

Our observation time is around 50\,ks that corresponds to $\sim$15\,ks or 4 hr in the rest frame of the source. As we do not see any variability, this implies that the BC of BLR photons is stable over this timescale. In fact, \cite{Celotti07} showed that the intensity of the BC spectrum from scattered BLR photons increases with time as the jet is accelerating and remains constant once the jet has reached its maximum bulk Lorentz factor. Given $kT_{\rm BB}$, we can estimate the factor $\delta\Gamma_{\rm bulk}$, where $\delta = \Gamma_{\rm bulk}^{-1}(1-\beta \cos \theta_{\rm V})^{-1}$ is the relativistic Doppler factor and $\theta_{\rm V}$ is the angle between the observer and the jet axis. Following \cite{Celotti07}, we assume that the BLR emission can be approximated by a blackbody spectrum peaking at the energy of the Ly$\alpha$, $h\nu_{{\rm Ly}\alpha} = 2.8 kT_{\rm BLR} = 13.6$\,eV. The observed temperature of the BC component (in the rest frame of the source) is then $kT_{\rm BLR,\,obs} = \delta \Gamma_{\rm bulk} kT_{\rm BLR}$, from which $\delta \Gamma_{\rm bulk} = 136$. One may also derive the expressions of $\theta_{\rm V}$ and the apparent superluminal speed $\beta_{\rm a}$ as a function of $\Gamma_{\rm bulk}$. For $\Gamma_{\rm bulk} = \delta = 11.7$, we find $\theta_{\rm V} \simeq 5^\circ$ and $\beta_{\rm a} \simeq 11.6$. Note that even if BC requires the presence of cold leptons, the jet cannot be pair dominated since in this case the plasma will suffer strong Compton drag and will be significantly decelerated \citep{Ghisellini10}. Alternatively, the jet could be magnetically dominated and proton free, but we consider this  possibility unlikely by noting that the SED of this source\footnote{The SED including all archival data can be built using the tool at \url{https://tools.asdc.asi.it/}.} is very similar to that of a typical FSRQ, for which \cite{Celotti08} model the high-energy spectrum as inverse Compton emission. The X-ray spectrum of FSRQ 4C+25.05 is rising (in $\nu F_\nu$) to higher energy, and the luminosity is already similar to that in the lower-energy synchrotron component. If the X-rays are due to inverse Compton scattering, the Compton peak dominates the bolometric electromagnetic output, and the magnetic-field component can only marginally contribute to the jet energy density.

BC emission is expected to be present in the X-ray spectra of all blazars. However, it has not been clearly detected until now. This can be mainly due to the presence of strong SSC and EC emission diluting any other emission features. In powerful FSRQs, the SSC is relatively weak, and the EC spectrum is exceptionally hard. These sources could be therefore the best candidates to search for BC signatures. We note that excesses similar to the one that we have found at soft energies have been reported is several blazars at redshifts below one \citep[e.g.][]{Piconcelli05, Sambruna06,Kataoka08,Derosa08}, and modelled with a blackbody component with $kT_{\rm BB}\sim 0.1$--0.2\,keV. However, this temperature range is lower than expected for the BC process, leading to low values of $\Gamma_{\rm bulk}\sim 5$--6. This makes 4C+25.05 an exceptional source where a softening in the X-ray spectrum has been detected for the first time, to our knowledge, above $z=1$, contrary to the soft X-ray flattening that has been observed up to $z=4.72$ \citep[e.g.][]{Worsley06}. In our case, the hard X-ray photon index obtained by applying a BB+PL model is only slightly flatter than the one obtained by fitting a simple PL ($\Delta \Gamma \sim 0.15$). This difference would be negligible for low S/N at high energies. The failure in detecting BC, in the past, could be due to the possible overestimate of the slope of the hard X-ray continuum. Our results show that simultaneous {\it XMM-Newton} and {\it NuSTAR} observations, providing high S/N spectra over the observed 0.3--30\,keV band, are capable of revealing the presence of the elusive BC feature at soft energies. We finally note that BC is expected to be a transient feature. Hence, a further and longer monitoring of the source, catching it probably in other spectral states, would be needed in order to confirm this scenario. Moreover, it could be worth revisiting the model of BC effects, based on more general assumptions and geometries, including a disc-like structure for the BLR (similar to the one observed in radio-quiet AGN) instead of the semi-spherical shell geometry considered by \cite{Celotti07}.

\section*{Acknowledgements}
EN is supported by the EU through the Horizon 2020 Marie Sk\l odowska-Curie grant agreement no. 664931. EB is supported by the EU through the Horizon 2020 Marie Sk\l odowska-Curie grant agreement no. 655324. The results presented in this paper are based on data obtained with the {\it NuSTAR} mission, a project led by the California Institute of Technology, managed by the Jet Propulsion Laboratory, and funded by NASA; {\it XMM-Newton}, an ESA science mission with instruments and contributions directly funded by ESA member states and NASA. This research has made use of the NuSTAR Data Analysis Software (NuSTARDAS), jointly developed by the ASI Science Data Center (Italy) and the California Institute of Technology (USA). The figures were generated using {\tt matplotlib} \citep{Hunter07}, a {\tt PYTHON} library for publication of quality graphics. The MCMC results were presented using the open source code {\tt corner.py} \citep{corner}.

Facilities: {\it NuSTAR},  {\it XMM-Newton}.

\bibliographystyle{mnras}
\bibliography{ek-4Cref} 

\bsp	
\label{lastpage}
\end{document}